\documentclass[preprint,11pt, a4paper]{elsarticle}

\usepackage{amssymb}
\usepackage{amsmath}
\usepackage{amsfonts}
\usepackage{graphicx,subcaption,float}
\usepackage{tikz}
\usepackage{wasysym}
\usepackage[mode=text, detect-none]{siunitx}
\usepackage{caption}
\usepackage{varwidth}
\usepackage{textcomp}
\usepackage{hyperref}

\usepackage[T1]{fontenc}
\usepackage{fouriernc}

\usepackage{lineno}

\journal{Software Impacts}
\makeatletter
\def\ps@pprintTitle{%
	\let\@oddhead\@empty
	\let\@evenhead\@empty
	\def\@oddfoot{\reset@font\hfil\thepage\hfil}
	\let\@evenfoot\@oddfoot
}
\makeatother

\usepackage{geometry}
\begin{document}

\begin{frontmatter}



\title{HEWES: Heisenberg-Euler Weak-Field Expansion Simulator}


\author{Andreas Lindner\corref{cor1}}
\ead{and.lindner@physik.uni-muenchen.de}
\author{Baris \"Olmez\corref{cor2}}
\ead{b.oelmez@physik.uni-muenchen.de}
\author{Hartmut Ruhl\corref{cor2}}
\ead{hartmut.ruhl@physik.uni-muenchen.de}
\address{Arnold Sommerfeld Center for Theoretical Physics, Ludwig-Maximilians-Universit\"at M\"unchen\\Theresienstr. 37, D-80333 M\"unchen, Germany}

\begin{abstract}
Vacuum polarization, a key prediction of quantum theory, can cause a variety of intriguing phenomena that can be triggered by high-intensity laser pulses.
The Heisenberg-Euler theory of the quantum vacuum supplements Maxwell's theory of electromagnetism with nonlinear photon-photon interactions mediated by vacuum fluctuations.
This work presents a numerical solver for the leading weak-field Heisenberg-Euler corrections.
The present code implementation reaches an accuracy of order thirteen in the numerical scheme and takes into account up to six-photon interactions.
Since theoretical approaches are limited to approximations and the experimental requirements for signal detection are high, the need for support from the numerical side is apparent.

\end{abstract}


\end{frontmatter}

\vspace{1cm}
\begin{figure}[H]
	\centering
	\includegraphics[width=\textwidth]{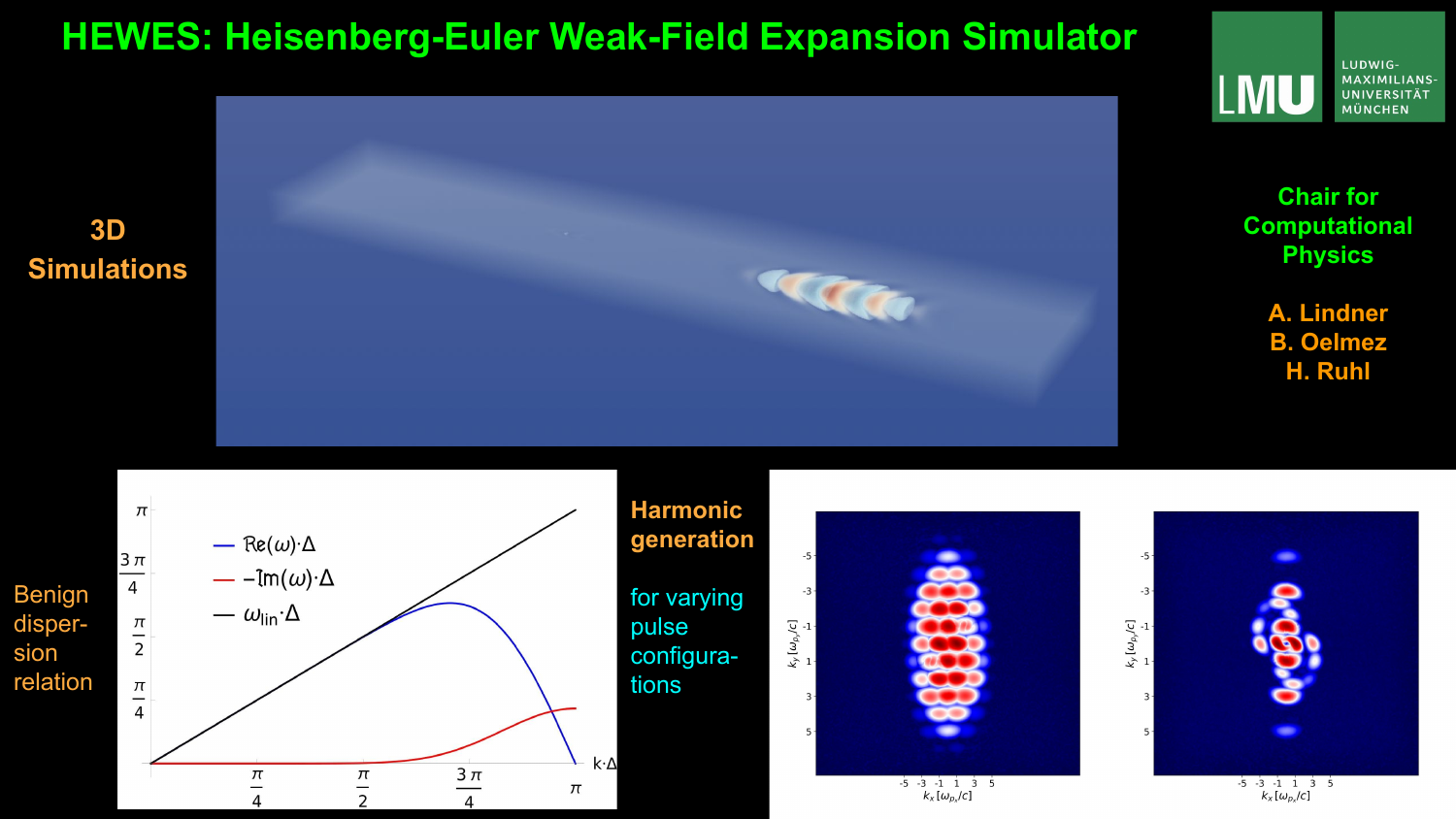}
\end{figure}
\vspace{1cm}

\section*{Highlights}
\begin{itemize}
	\item A universal 13\textsuperscript{th} order of accuracy numerical scheme with a nonphysical modes filter
	\item Scalability on distributed computing systems
	\item Universality with respect to pulse configurations in contrast to analytical treatments
	\item Versatility in the context of high-precision experiments
	\item Inclusion of up to six-photon processes in the Heisenberg-Euler weak-field expansion
\end{itemize}


\section{Motivation and significance}

The code presented in this work solves the nonlinear equations of the low-energy effective theory of quantum electrodynamics (QED) by Heisenberg and Euler \cite{HeisenbergEuler1936}.
The theory describes quantum vacuum polarization in the form of virtual electron-positron pairs capable of mediating photon-photon interactions.
The QED vacuum can hence be understood as a nonlinear, polarizable medium.
Nonlinear optical effects should in principle be observable.

The Heisenberg-Euler theory can be expressed purely in terms of radiation fields in a weak-field expansion that extends Maxwell's linear theory of electromagnetism with nonlinear interactions.
The violations of the superposition principle for electromagnetic fields thereby become noticeable only at high intensities of the involved light sources.
Experimental and financial hurdles for their detection are thus very high.
Facilities need to be equipped with high-intensity laser pulses and likewise ultra sensitive detectors.
Advanced numerical frameworks are therefore relevant to simulate future QED experiments.

QED is traditionally tested with collider experiments in the high-energy, low-intensity regime.
Approaching the theory via the quantum vacuum tackles the low-energy, high-intensity regime instead.
Due to the lack of sensitive high-intensity experiments, this regime has not been accessible thus far.

Nevertheless, it might serve as a portal towards new physics as well.
The particle-antiparticle fluctuations in the vacuum consist of all existing particles.
Hence, the influence of yet unknown particle fields might be inscribed in quantum vacuum signatures \cite{Gies2008,Karbsteinetal2019}.
A currently hot topic is the anomalously large magnetic moment of the muon,
where quantum vacuum polarization is suspected to be a key factor contribution \cite{Aoyamaetal2020,Abietal2021,Borsanyietal2021}.

A promising approach to the detection of nonlinear \textit{all-optical} effects is the investigation of the asymptotic dynamics of probe photons after passing a strong electromagnetic field region in the form of a high-intensity laser.
Probe photons traversing a strong field region can indirectly sense the applied pump field via the quantum fluctuations that couple to both probe and pump fields \cite{KarbsteinSundquist2016}.
Due to the nonlinear interaction with the power pulse, probe photons scatter in such a way that their dynamics or polarization distinguishes them from the latter.
The properties and dynamics of the quantum vacuum, in turn, are hereby encoded into the probe photons.
This class of quantum vacuum experiments, where one electromagnetic field drives the nonlinear effect while the other carries its signature, is therefore called all-optical.
These experiments at the high-intensity frontier promise unprecedented detail in the study of the quantum vacuum in the near future \cite{MarklundLundin2009}.
For reviews, see \cite{DittrichReuter1985,DittrichGies2000,MarklundLundin2009,Dunne2009,HeinzlIlderton2009,DiPiazzaetal2012,Dunne2012,KingHeinzl2016,Karbstein2016b,Inadaetal2017}.

\section{Theoretical background}

\begin{figure}
	\centering
	\includegraphics[width=\textwidth]{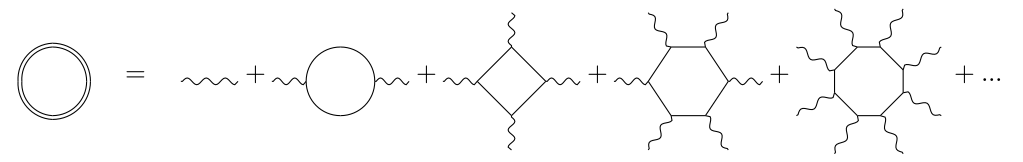}
	\caption[QED vacuum weak-field expansion]{Depiction of the weak-field expansion of the QED vacuum. 
		The double-lined loop to the left represents the full Lagrangian, while the diagrams to the right represent the linear Maxwell theory, the one-loop correction to it, and the four-, six-, and eight-photon contributions due to the Heisenberg-Euler weak-field expansion.
		\label{fig:Diagramme}}
\end{figure}

The Heisenberg-Euler Lagrangian \cite{Schwinger1951} can be expanded in a weak-field form, where "weak" means field strengths below the \textit{critical} field strength of the QED vacuum $ E_{\textrm{cr}}= {m_e^2 c^3}/(e\hbar) =\SI{1.323e18}{\volt \per \metre}$, where $ e $ is the charge of the electron, $ m_e $ its mass, $ c $ the vacuum speed of light, and $ \hbar $ the reduced Planck constant.
Electric fields with this intensity open up a new regime of strong-field QED, where the properties of the vacuum are substantially different \cite{Sauter1931,Schwinger1951,Heinzl2012}.
The nonlinear corrections to Maxwell's theory of electromagnetism at the one-loop level due to the weak-field expansion of the Heisenberg-Euler theory can be depicted as in Figure \ref{fig:Diagramme}.

Notably, pair production is exponentially suppressed in the complete Heisenberg-Euler Lagrangian for below-critical field strengths.
Therefore, the properties of the nonlinear vacuum in the weak-field limit can be described exclusively by radiation fields.

\section{Description}

\subsection{Overview of the implementation}

\begin{figure}
	\centering
	\begin{subfigure}{.45\textwidth}
		\includegraphics[width=\linewidth]{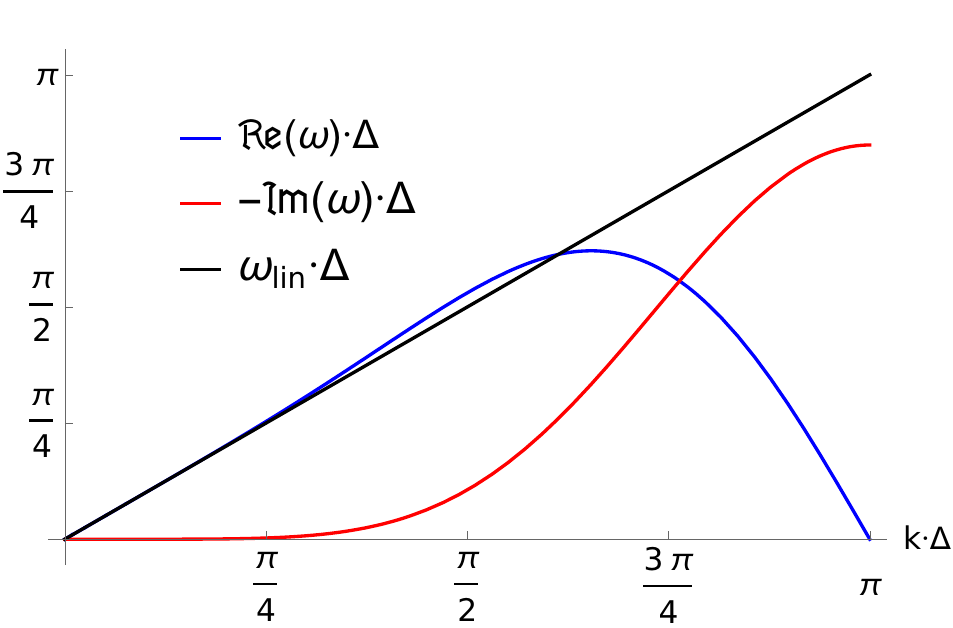}
	\end{subfigure}\qquad\quad	
	\begin{subfigure}{.45\textwidth}
		\includegraphics[width=\linewidth]{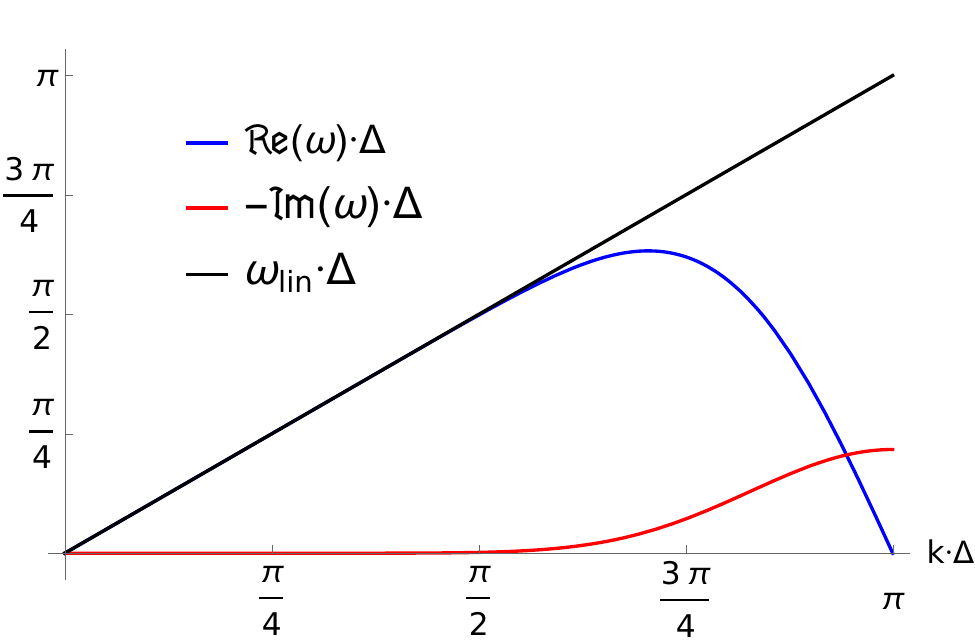}
	\end{subfigure}
	\caption[Dispersion relation]{Dispersion relations of the applied numerical scheme at order four (\textit{left}) and order thirteen (\textit{right}).
		The black lines represent the real vacuum dispersion relation between frequency $\omega$ and wavenumber $k$, in particular $\omega = c \cdot k$ by choosing $c=1$.
		$\Delta$ is the grid spacing, the physical distance between lattice points.
		$\omega$ obtains an imaginary part in the scheme \cite{Lindneretal2023}.
		The real parts start in the vicinity of the black line and deviate from it for shorter wavelengths or smaller grid resolutions.
		There is less deviation at higher orders.
		At the wavelength corresponding to the Nyquist frequency, given by $k \cdot \Delta = \pi$, the real part of $\omega$ is zero for all orders.
		The imaginary part of $\omega$ starts with values close to zero for large wavelengths/high grid resolutions and decreases until it reaches a minimum at the Nyquist frequency.
		The negative imaginary part in this scheme has the positive effect to quickly annihilate those modes that deviate strongly from the vacuum dispersion relation.
		At higher orders, the damping effect of the imaginary part is deferred to shorter wavelengths and is overall smaller.}
\label{fig:dispersion}
\end{figure}

The simulation code presented in this work solves the Maxwell equations extended by up to six-photon interactions due to the weak-field expansion of the Heisenberg-Euler theory.
The algorithm for solving the modified Maxwell equations is discussed to some extent in \cite{Lindneretal2023}.
Depending on the chosen order of the numerical scheme, it possesses an almost linear vacuum dispersion relation even for smaller wavelengths.
This allows the use of comparatively small grids.
Moreover, an imaginary part in the dispersion relation annihilates nonphysical modes, see Figure \ref{fig:dispersion}.

Partial differential equations are turned into ordinary differential equations (ODEs) by means of a finite difference scheme on a spatially discrete lattice.
The devised scheme and the implementation work in one to three spatial dimensions plus time.
Hyperparameter values at the user's option determine the overall accuracy.
These are the order of the finite difference scheme -- ranging from one to thirteen -- the lattice resolution, and the error tolerances of the employed ODE solver.
The \textit{SUNDIALS} \cite{Hindmarshetal2005,Gardneretal2022} package is used for solving the arising system of nonlinear differential equations.
Explicitly, the \textit{CVODE} solver \cite{Hindmarshetal2021} from the \textit{SUNDIALS} family of solvers is employed, configured to use the Adams method (Adams-Moulton formula) \cite{Haireretal1993} in conjunction with a fixed-point iteration.

Cluster computer communication for large-scale simulations with distributed memory is guided by the Message Passing Interface (\textit{MPI}) \cite{Clarkeetal1994} on a virtual Cartesian topology, if \textit{MPI} is enabled.
The software is coded in \textit{C++} with features up to the \textit{C++20} standard.
Use of \textit{OpenMP} \cite{Mattson2019} is optional to enforce more vectorization and enable multi-threading.
The latter is useful for performance only at a scale of about 1000 compute cores.

\subsection{Usage}
\label{sec:usage}

There is full control over all high-level simulation settings with command line arguments.
A list of input parameters along with descriptions is given in the README file.
\textit{Bash} and \textit{Windows Powershell} example run scripts are provided in the repository.

Simulations can be performed on a one to three dimensional lattice.
It can be decided whether to simulate in the linear Maxwell vacuum, the linear vacuum plus four-photon interactions, the linear vacuum plus six-photon interactions, or the linear vacuum plus four- and six-photon interactions.

The accuracy of the \textit{CVODE} integrator and the order of the numerical scheme can be set.
The number of steps performed in the total physical propagation time is another parameter that is crucial for accuracy.

Implementations of Gaussian laser pulses are available and their parameters can be configured.

The output consists of all electromagnetic field components, namely $E_x \, , \ E_y \, , \ E_z \, , \ B_x \, , \ B_y$, and $B_z $, at every grid point.
The output format can be comma-separated-values or binary and it can be chosen after how many steps the field data are written to disk.
\textit{Python} modules are provided to read the electromagnetic field components into \textit{NumPy} arrays for then ensuing analysis.
The configuration of the grid and the decomposition into \textit{MPI} processes becomes relevant in higher dimensions.

Example analyses are provided in the code repository and in a \textit{Mendeley Data} repository \cite{Lindner2022}.

\subsection{Note on resource occupation}

The computational load mostly depends on the grid size and resolution.
The order of accuracy of the numerical scheme and \textit{CVODE} are rather secondary, except for simulations running on many processing units. 
There, the communication load plays a major role, which in turn depends on the order of the numerical scheme.
This is because the order of the scheme determines how many neighboring grid points are taken into account for the finite difference derivatives.
To fully exploit the beneficial dispersion properties, it is nevertheless recommended to use the highest available order (thirteen) of the scheme.

For the absolute and relative error tolerances of the \textit{CVODE} solver $ 10^{-12}$ or lower are good choices.
The Adams method is preset to use the highest available order (twelve).
The step size should not be larger than $2 \mu m$.

Standard simulations in 1D can easily be run on a modern notebook within seconds.
The output size per step is less than a megabyte.
Simulations in 2D running on about one million grid points are still feasible for a personal computer and take only a couple of minutes.
The output size per step is in the range of some dozen megabytes.
Sensible simulations in 3D require large memory resources and therefore need to be run on distributed systems.
This implies an increased communication load.
Even hundreds or thousands of cores can be occupied for many hours or days.
This forms a practical limit to the grid resolution \cite{Lindneretal2023}.
The output size can amount to hundreds of gigabytes for just a single time step on high-resolution grids.

A 3D simulation to produce the results shown in Figure \ref{fig:3d} can be performed on a grid with $1400\times 1400 \times 200$ points in less than four hours on about 400 compute cores.
The output size amounts to nearly 20 gigabytes for one time step.
A weak scaling test for such 3D simulations is demonstrated in \cite{Lindneretal2021}.
The computational load scales linearly with the grid size and the simulation time varies only slightly when the load is equally distributed on up to about 1000 cores.

\section{Impact}

\begin{figure}
	\centering
	\begin{subfigure}{.48\textwidth}
		\centering
		\includegraphics[width=\linewidth]{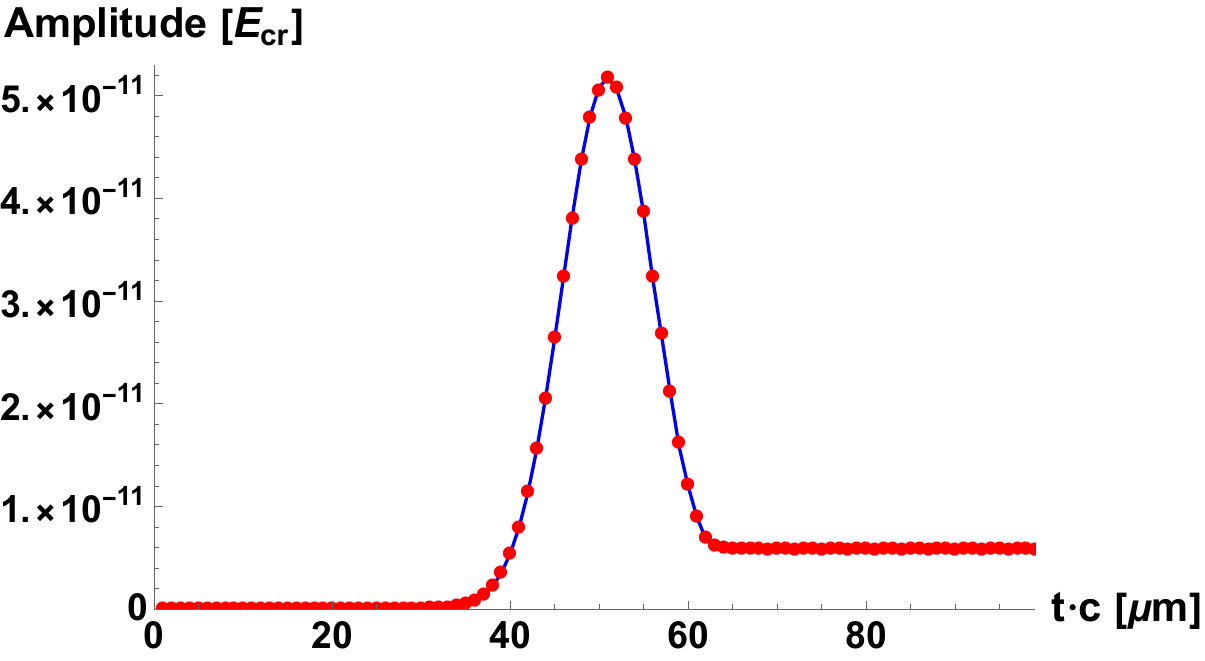}
		\if false
		\caption[Second harmonic time evolution]{Time evolution of the amplitude of a second harmonic caused by nonlinear effects.
			This high-harmonic is caused by a probe pulse colliding with a zero-frequency background pump pulse.
			The simulation results of \cite{Lindneretal2023} (\textit{red dots}) completely agree with the analytical approximation obtained in \cite{Kingetal2014} (\textit{blue line}).
			\label{fig:harmonic_with_ana}}
		\fi
	\end{subfigure}\qquad
	\begin{subfigure}{.41\textwidth}
		\centering
		\includegraphics[width=\linewidth]{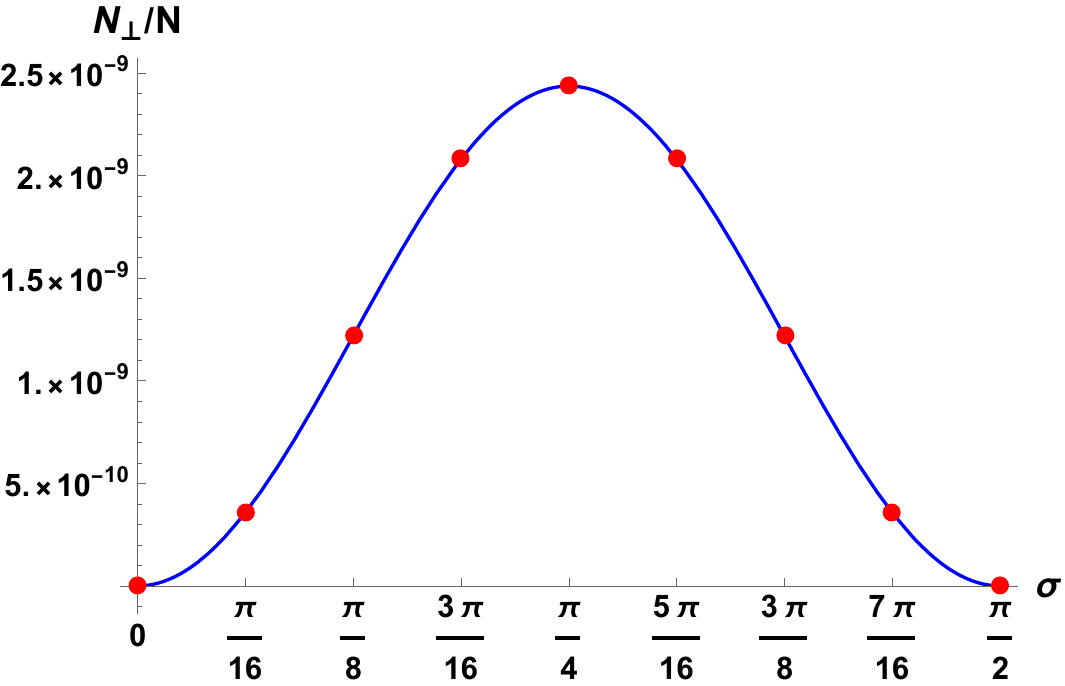}
		\if false
		\caption[Relative angle dependency in polarization rotation]{Polarization flipping ratio dependency on the relative polarization angle of probe and pump pulse.
			The values obtained via simulations in \cite{Lindneretal2023} (\textit{red dots}) are in perfect agreement with the analytical solution obtained in \cite{Dinuetal2014} (\textit{blue line}).
			\label{fig:pol_flip_s_scaling}}
		\fi
	\end{subfigure}
	\caption[1D simulations with analytical benchmarks]{Simulations in one spatial dimension with analytical benchmarks.
\textit{Left:} Time evolution of the amplitude of a second harmonic due to nonlinear vacuum effects.
This high-harmonic is caused by a probe pulse colliding with a zero-frequency background pump pulse.
The simulation results of \cite{Lindneretal2023} (\textit{red dots}) completely agree with the analytical approximation obtained in \cite{Kingetal2014} (\textit{blue line}).
\textit{Right:} Polarization flipping ratio ($N\perp/N$) dependency on the relative polarization angle of probe and pump pulse ($\sigma$).
The values obtained via simulations in \cite{Lindneretal2023} (\textit{red dots}) are in perfect agreement with the analytical solution obtained in \cite{Dinuetal2014} (\textit{blue line}).}
	\label{fig:1d_sims}
\end{figure}

\begin{figure}
	\centering
	\begin{subfigure}{.49\textwidth}
		\centering
		\includegraphics[width=\linewidth]{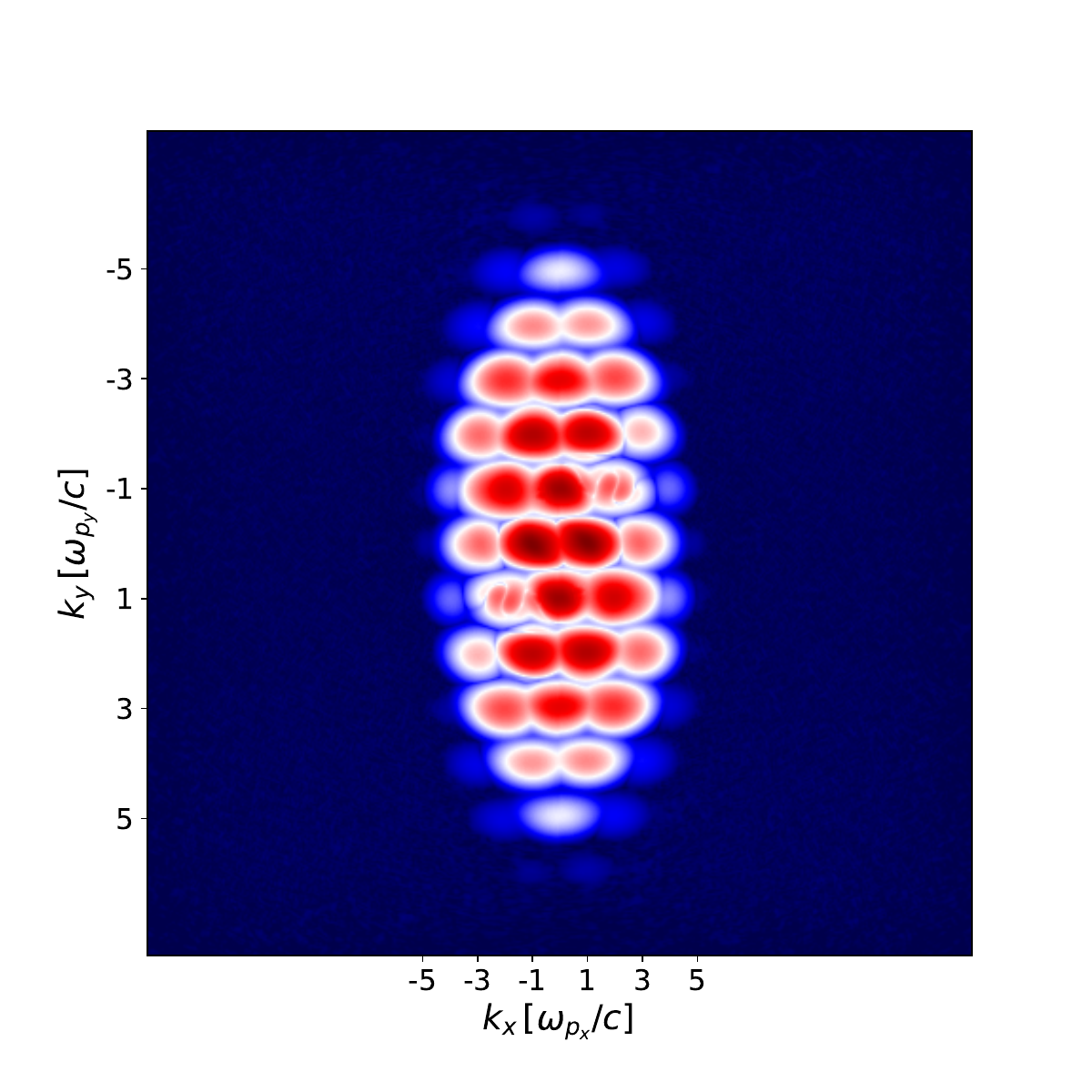}
	\end{subfigure}
	\begin{subfigure}{.49\textwidth}
		\centering
		\includegraphics[width=\linewidth]{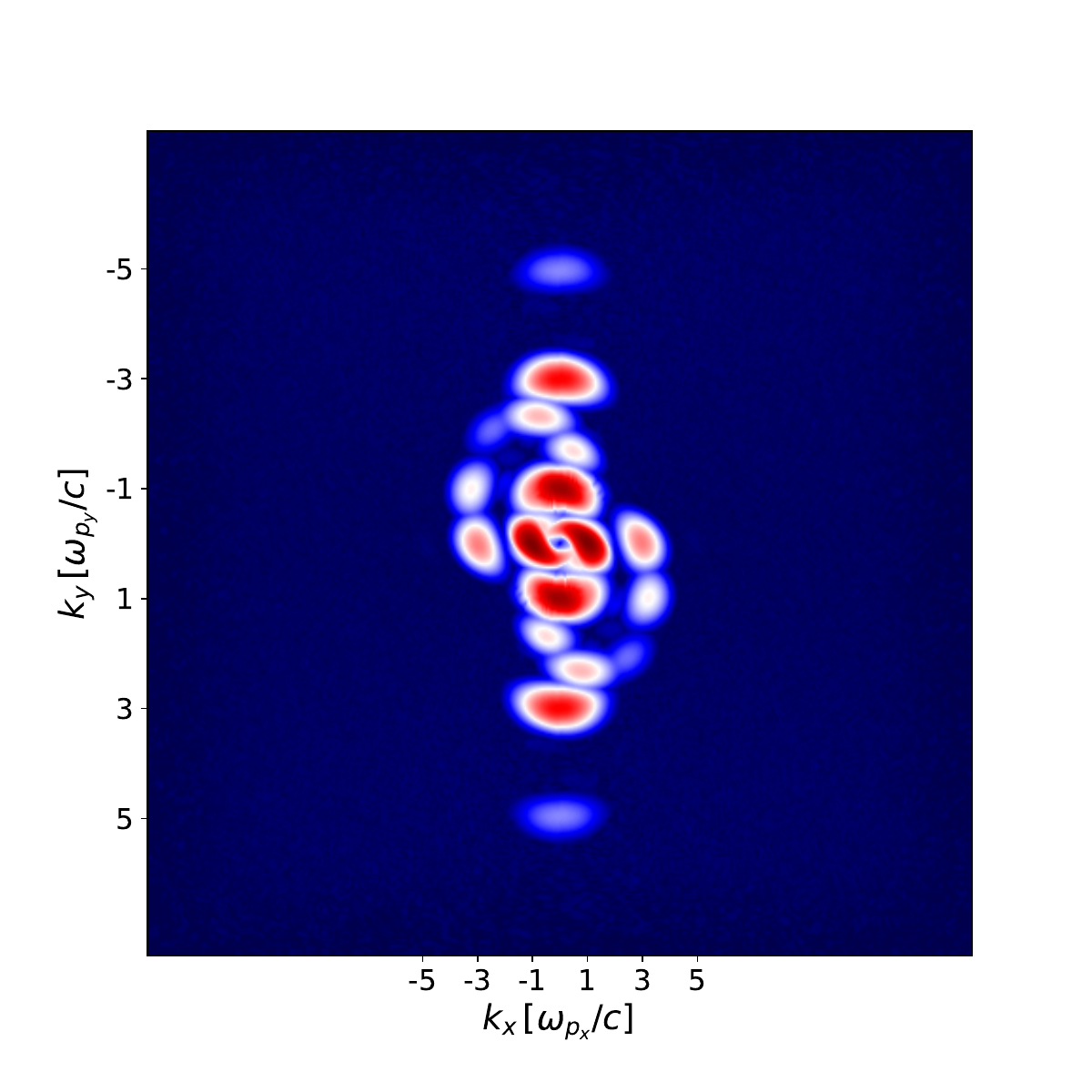}
	\end{subfigure}
	\caption[Harmonic generation in 2D with different frequency pulses]{Generation of higher harmonics in two spatial dimensions.
		Solely contributions from nonlinear vacuum effects are illustrated.
		In this case, perpendicularly colliding Gaussian pulses with different intensities and wavelengths generate a rich frequency spectrum during the pulse overlap position (\textit{left}) and asymptotic harmonics persisting after the pulses have separated (\textit{right}).
		The wavenumbers $ k $ denoting the axes are given in terms of the frequencies $ \omega $ of the involved pulses.}
	\label{fig:corn_yinyang}
\end{figure}

\begin{figure}
	\centering
	\begin{subfigure}{.49\textwidth}
		\centering
		\includegraphics[width=\linewidth]{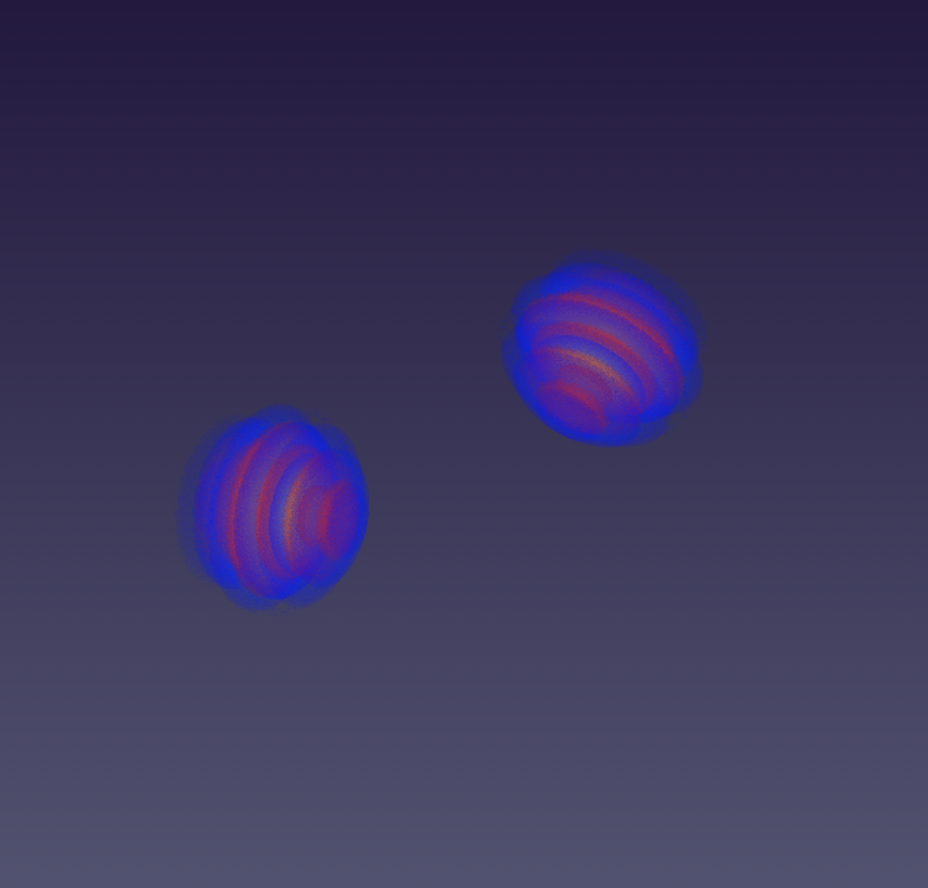}
	\end{subfigure}
	\begin{subfigure}{.49\textwidth}
		\centering
		\includegraphics[width=\linewidth]{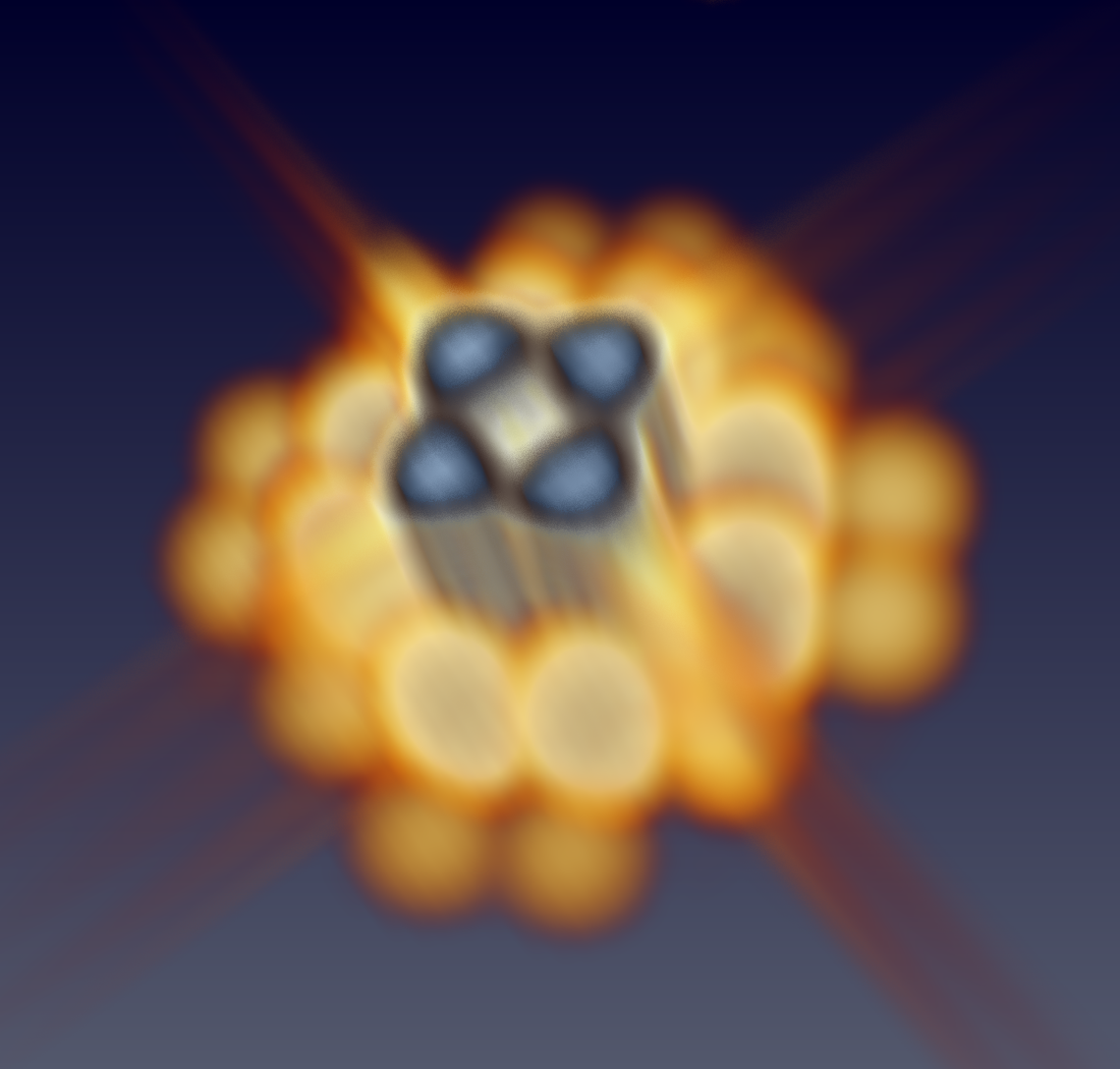}
	\end{subfigure}
	\caption[Harmonic generation in 3D]{Generation of higher harmonics in three spatial dimensions.
		\textit{Left:} Initial configuration of the pulses in position space.
		\textit{Right:} Frequency spectrum at the overlap position of the pulses, where a rich spectrum is observable again.}
	\label{fig:3d}
\end{figure}

Originating from a PhD project \cite{Pons2018}, the code has been further developed and in \cite{Lindneretal2023} various effects of nonlinear quantum vacuum theory have been demonstrated with simulations.
The current focus of the project lies on vacuum birefringence \cite{Toll1952,BaierBreitenlohner1967,Bialynicka-BirulaBialynicki-Birula1970} and high-harmonic generation \cite{BhartiaValluri1978,ValluriBhartia1980,Bialynicka-Birula1981}.
In these effects, the effective photon-photon interactions during the collision of laser pulses cause a rotation of the polarization direction or the arising of higher-frequency modes, respectively.
A selection of results of simulation in 1D, 2D, and 3D is depicted in Figures \ref{fig:1d_sims}, \ref{fig:corn_yinyang}, and \ref{fig:3d}.

Some simulations in one spatial dimension can be cross-checked against analytical results and are in perfect agreement with the latter.
On the other hand, the simulation results in higher dimensions have no direct analytical counterparts to compare with.
A comparable accuracy can be expected.

Analytical methods rely on simplifying assumptions and are therefore most of the time limited to special scenarios.
Feasible calculations are constrained to simple configurations and arrangements of the involved laser pulses, or neglect important properties of the quantum vacuum.
Any such approximation in turn limits the accuracy of predictions and the precision with which theory can be tested \cite{Blinneetal2019a}.

\begin{figure}
	\centering
	\includegraphics[width=\linewidth]{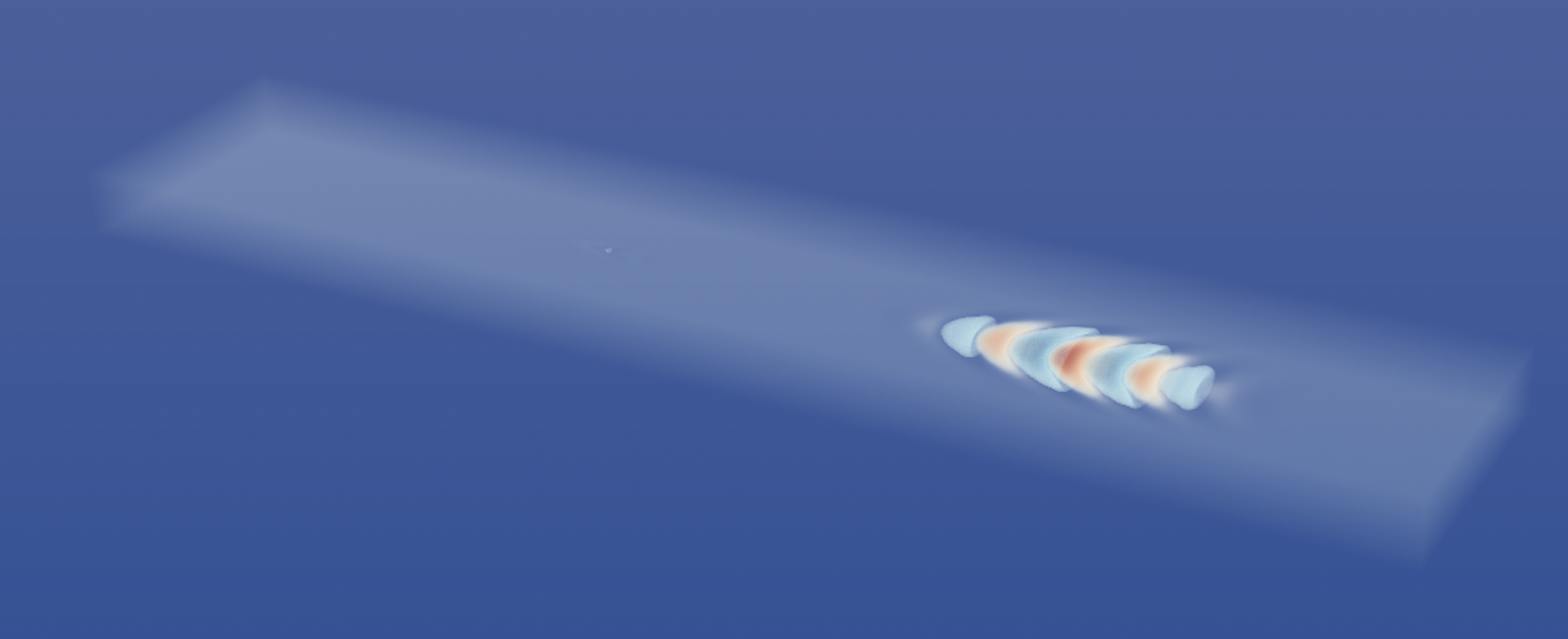}
	\caption[Probe-pump setup in 3D]{Probe-pump collision in a flat-box shaped simulation space in order to detect a polarization rotation of the probe pulse.
	The probe pulse on the left is tiny in comparison to the strong pump on the right.}
	\label{fig:tube}
\end{figure}

The numerical solver is agnostic to the specific configuration and yields the complete picture of the quantum vacuum, within the limits of the  Heisenberg-Euler weak-field approximation.
While theoretical calculations confine themselves to single effects, the solver takes into account the whole picture of the nonlinear vacuum dynamics, including back-reactions to the radiation fields.
Thereby, the solver is capable of describing the temporal evolution of the processes, while theoretical approaches oftentimes concentrate only on asymptotic values.
A promising configuration for the detection of a nonlinear vacuum response is given by the prominent probe-pump laser pulse collision setup shown in Figure \ref{fig:tube}.
Predictions for light-by-light scattering in general should be possible in all conceivable interaction scenarios between probe and pump with the help of the solver \cite{Lindneretal2023}.

Simulation results in higher dimensions extend the horizon to yet undiscovered terrain that is particularly relevant for experiments.
The exploration of parameter regimes and the estimation of expected signals that should be noticeable in experiments will be supported by numerical tools.
Moreover, only computer-driven approaches are flexible enough to guide the development of experimental constructions and configurations in the research area of strong-field QED.
A shift in perspective to accompany the numerous analytical considerations with versatile numerical solutions is apparent.

There have been other approaches in this research area, put forward in \cite{Giesetal2018,Blinneetal2019a} and  \cite{Grismayeretal2021}.
The one discussed in the present paper stands out with a very high order of accuracy of the numerical scheme and the inclusion of six-photon processes, and is thus extremely precise.
At the heart of the algorithm lies the dispersion relation that ensures stability throughout the frequency spectrum and on top creates an imaginary part that annihilates nonphysical modes \cite{Lindneretal2023}.

On the numerical side, ideas are being developed
in order to overcome the obstacle of extremely large 3D grids.
One promising, ongoing, and important project is on multi-scale simulation capability.
This can be achieved by adaptive data structures and integrators combined with novel machine learning concepts.

\section*{Acknowledgments}

The software development is part of the German Research Foundation (DFG) Research Unit \href{http://www.quantumvacuum.org/index.html}{FOR 2783} "Probing the Quantum Vacuum at the High-Intensity Frontier" and has been funded under the Grant Nos. \href{https://gepris.dfg.de/gepris/projekt/416611371}{416611371}; \href{https://gepris.dfg.de/gepris/projekt/416607684}{416607684}.

Large parts of the computations during the production and verification process have been performed on the KSC cluster computing system of the Arnold Sommerfeld Center (ASC) for Theoretical Physics at LMU Munich, hosted at the Leibniz-Rechenzentrum (LRZ) in Garching and funded by the German Research Foundation under Grant No. \href{https://gepris.dfg.de/gepris/projekt/409562408}{409562408}.

The hospitality of the Arnold Sommerfeld Center is acknowledged.

\section*{Data and code availability}
The raw simulation data used for this work amount to more than 100 GB in size.
They are archived on servers of the Arnold Sommerfeld Center (ASC) for Theoretical Physics in Munich, hosted by the Leibniz-Rechenzentrum (LRZ), in compliance with the regulations of the German Research Foundation (DFG).
There is a reproducible code capsule published on \textit{Code Ocean} \cite{Lindner2023} and a \textit{Mendeley Data} repository containing extra and supplementary materials \cite{Lindner2022}.

\section*{Code metadata}

\begin{table}[!h]
	\centering
	\begin{tabular}{|p{6.5cm}|p{6.5cm}|}
		\hline
		Current code version & v0.2.5 \\
		\hline
		Link to code/repository  & \href{https://gitlab.physik.uni-muenchen.de/ls-ruhl/hewes}{https:\slash\slash gitlab.physik.uni-muenchen.de\slash ls-ruhl\slash hewes} \\
		\hline
		Link to Reproducible Capsule & \href{https://codeocean.com/capsule/3187285/tree}
		{https:\slash\slash codeocean.com\slash capsule\slash 3187285\slash tree} \\
		\hline
		Legal Code License   & \href{https://gitlab.physik.uni-muenchen.de/ls-ruhl/hewes/-/blob/main/LICENSE}{BSD 3-Clause License} \\
		\hline
		Code versioning system used & git \\
		\hline
		Software code languages, tools, and services used & C++20, (MPI-3.1, OpenMP v4.5) \\
		\hline
		Compilation requirements, operating environments \& dependencies & CMake $\geq \,$ v3.21
		\\
		\hline
		Link to code reference &
		\href{https://gitlab.physik.uni-muenchen.de/ls-ruhl/hewes/-/blob/main/README.md}{https:\slash\slash gitlab.physik.uni-muenchen.de\slash ls-ruhl\slash hewes\slash -\slash blob\slash main\slash README.md}\\ &  \href{https://gitlab.physik.uni-muenchen.de/ls-ruhl/hewes/-/blob/main/docs/ref.pdf}{https:\slash\slash gitlab.physik.uni-muenchen.de\slash ls-ruhl\slash hewes\slash -\slash blob\slash main\slash docs\slash ref.pdf} \\
		\hline
		Support email for questions & \href{mailto:and.lindner@physik.uni-muenchen.de}{and.lindner@physik.uni-muenchen.de} \\
		\hline
	\end{tabular}
\end{table}

\bibliographystyle{elsarticle-num}
\biboptions{sort&compress}  
\bibliography{../../refs}

\end{document}